\documentclass[preprint]{vgtc}               % preprint
\ifpdf%                                % if we use pdflatex
  \pdfoutput=1\relax                   % create PDFs from pdfLaTeX
  \usepackage{graphicx}                % allow us to embed graphics files
  \DeclareGraphicsExtensions{.pdf,.png,.jpg,.jpeg} % for pdflatex we expect .pdf, .png, or .jpg files
\else%                                 % else we use pure latex
  \usepackage{graphicx}                % allow us to embed graphics files
  \DeclareGraphicsExtensions{.eps}     % for pure latex we expect eps files
\fi%

\graphicspath{{figures/}{pictures/}{images/}{./}} % where to search for the images

\usepackage{microtype}                 % use micro-typography (slightly more compact, better to read)
\PassOptionsToPackage{warn}{textcomp}  % to address font issues with \textrightarrow
\usepackage{textcomp}                  % use better special symbols
\usepackage{mathptmx}                  % use matching math font
\usepackage{times}                     % we use Times as the main font
\usepackage{cite}                      % needed to automatically sort the references
\usepackage{tabu}                      % only used for the table example
\usepackage{booktabs}                  % only used for the table example
\usepackage[table]{xcolor}
\usepackage{graphics}
\usepackage{amssymb}

\definecolor{c1}{rgb}{0,0.3,1}
\definecolor{c2}{rgb}{1,0,0.0}
\definecolor{c3}{rgb}{0.16, 0.5, 0.0}
\definecolor{c4}{rgb}{0.2, 0.41, 0.65}

\newcommand{\newtechniquename}{\textit{dg2pix}\thinspace}
\onlineid{1001}

\vgtccategory{Research}

\vgtcinsertpkg

\title{dg2pix: Pixel-Based Visual Analysis of Dynamic Graphs}

\author{Eren Cakmak\thanks{e-mail: eren.cakmak@uni-konstanz.de}\\ %
        \scriptsize University of Konstanz %
\and Dominik J\"ackle\thanks{e-mail: dominikjaeckle@gmail.com}\\ %
     \scriptsize Independent Researcher%
\and Tobias Schreck\thanks{e-mail: tobias.schreck@cgv.tugraz.at}\\ %
     \scriptsize TU Graz %
\and Daniel Keim\thanks{e-mail: keim@uni-konstanz.de}\\ %
     \scriptsize University of Konstanz %
}

\teaser{
  \centering
  \includegraphics[width=.94\linewidth]{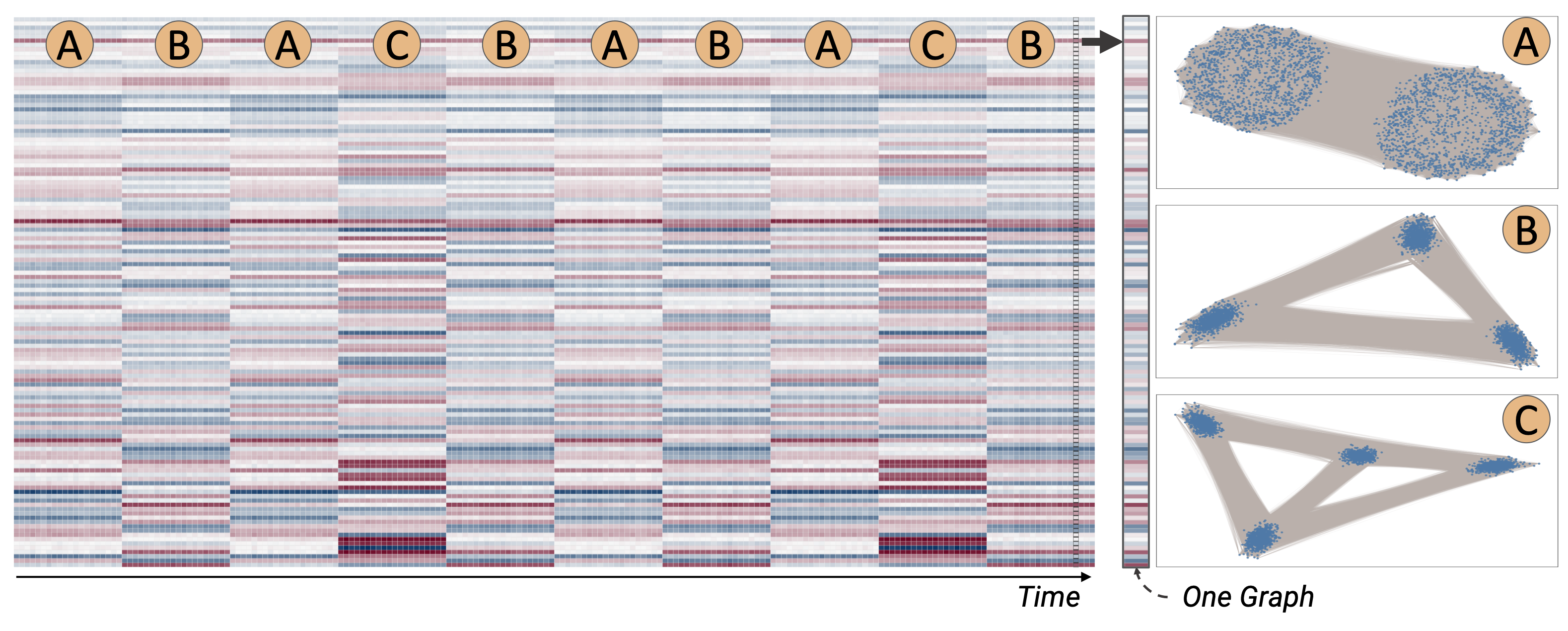}
  \vspace{-0.4cm}
  \caption{ % Why 
  \newtechniquename provides an overview of temporal and structural changes in dynamic graphs.
  The example presents a synthetic dynamic graph (200-time steps) using graph2vec~\cite{narayanan2017graph2vec}.
  The x-axis presents the temporal dimension, and the y-axis displays for each time step a graph embedding as a pixel-bar.
  The reoccurring states (A-C) have for each time step the same amount of nodes (2500) and edges (350000) with a different number of clusters.
  Each state (20-time steps) was generated with SBM~\cite{holland1983stochastic} with slight variations for the density of edges between clusters. 
  The graphs (A-C) display a sample graph for each state.
  \newtechniquename enables us to explore and identify temporal changes, outlier graphs, and reoccurring graph structures at multiple temporal scales. 
}
  \label{fig:teaser}
}

\abstract{
Presenting long sequences of dynamic graphs remains challenging due to the underlying large-scale and high-dimensional data.
We propose \newtechniquename, a novel pixel-based visualization technique, to visually explore temporal and structural properties in long sequences of large-scale graphs.
The approach consists of three main steps: (1) the multiscale modeling of the temporal dimension; 
(2) unsupervised graph embeddings to learn low-dimensional representations of the dynamic graph data; 
and (3) an interactive pixel-based visualization to simultaneously explore the evolving data at different temporal aggregation scales.
\newtechniquename provides a scalable overview of a dynamic graph, supports the exploration of long sequences of high-dimensional graph data, and enables the identification and comparison of similar temporal states.
We show the applicability of the technique to synthetic and real-world datasets, demonstrating that temporal patterns in dynamic graphs can be identified and interpreted over time. 
\newtechniquename contributes a suitable intermediate representation between node-link diagrams at the high detail end and matrix representations on the low detail end.
} % end of abstract
\CCScatlist{
   \CCScatTwelve{Human-centered computing}{Visualization}{Visualization techniques}{};
   \CCScatTwelve{Machine learning}{Learning paradigms}{Unsupervised learning}{}
}

\begin{document}


\firstsection{Introduction}

\maketitle

%% \section{Introduction} %for journal use above 

%%%%%%%%%%%%%%%%%%%%%%%%%%%%%%%%%%%%%%%%%%%%%%%%%%%%%%%%%%%%%%%%%%%%%%%%%%%
% INTRODUCTION
%%%%%%%%%%%%%%%%%%%%%%%%%%%%%%%%%%%%%%%%%%%%%%%%%%%%%%%%%%%%%%%%%%%%%%%%%%%
\label{sec:intro}
% FIRST PARAGRAPH
% - what is the main characteristic of the data with examples
% - what is the KEY TASK!
% - what is the main problem/challenge regarding the task
% Dynamic graph visualization
Dynamic graph visualizations are used in many real-world applications to present evolving relationships between entities, such as in social network analysis. 
% The task - where in the graph, when in time, and what has happened? 
A primary user task in such dynamic graph visualizations is to obtain an overview of the temporal dimension, for instance, to identify temporal states such as trends, outlier graphs, and similar graph structures over time~\cite{van2016reducing}.
% Challenges
However, visualizing large-scale dynamic graphs remains challenging as such visualization techniques have to present large amounts of evolving high-dimensional data in a readable and scalable manner~\cite{beck2017taxonomy}.

%%%
% SECOND PARAGRAPH
% - related approaches and what their limitations are - high level
% - what are the main challenges
% - what do we propose described in one single sentence
Visualization techniques for dynamic graphs can be distinguished by the following primary categories: animation and timeline visualization~\cite{beck2017taxonomy}.
% Problem
However, both categories do not scale due to a large number of nodes, edges, and time steps~\cite{beck2009towards}.
% High dimensionality 
Particularly, the evolving, potentially highly complex data may pose a significant challenge for the visual detection and traceability of changes in dynamic graph visualizations.
% Data abstraction
Therefore, previous approaches for the visual analysis of dynamic graph data often rely on dimensionality reduction methods to provide an overview of higher-level structures over time~\cite{van2016reducing}.
% Advantages 
Such dimensionality reduction methods reduce the complexity by embedding the evolving topological structures in low-dimensional space.
% to present the evolving properties of large-scale dynamic graphs.
% What 
While to date, some dynamic graph visualization techniques leverage dimensionality reduction methods (e.g., 2D embeddings~\cite{van2016reducing}), they still fail to provide a scalable overview of the structural changes as the approaches depend on the temporal analysis scale and the designed feature vector (e.g., graph metrics). 
% Therefore, we introduce a multiscale pixel-based visualization technique based on unsupervised learning methods to provide a scalable overview of changes in a dynamic graph.

%%
% THIRD PARAGRAPH
% - what do we propose - high level
% - what is the research question
% - how does our approach work, what do we apply
% - what are the advantages of our approach
% - application examples and use cases we show to demonstrate the usefulness
We propose \newtechniquename (dynamic graph to pixel-based visualization), a novel visualization technique for large-scale dynamic graphs based on unsupervised graph learning methods (e.g., graph2vec~\cite{narayanan2017graph2vec} or FGSD~\cite{verma2017hunt}). 
% Main goal  
The main goal is to provide a scalable overview of the temporal dimension and enable the initial exploration of the high-dimensional data to support the identification of temporal changes and similar temporal states.
% How 
The visualization technique consists of three main steps: multiscale temporal modeling, graph embeddings, and a pixel-based visualization. 
% Advantages 
The graph embedding reduces the dynamic graph to a low-dimensional representation (50-300 dimensions) and learns the similarity between graphs to capture the evolving topology of the high-dimensional data.
% The technique 
The compact visualization technique allows users to interactively adapt the temporal analysis scale and compare high-level as well as fine-grained structural changes.
% Evaluation 
We demonstrate the usefulness of our approach through two use cases to show how \newtechniquename can be utilized to identify temporal changes and states in dynamic graphs.

% Contributions
In summary, the contributions of this work are the following: 
(1) The novel \newtechniquename visualization technique, a time-scalable visual metaphor to reveal changes and similar temporal states in a dynamic graph;
(2) an interpretation strategy of visual patterns that users can examine in \newtechniquename;
and (3) an interactive prototype that allows exploring dynamic graphs at multiple scales.

\section{Related Work} \label{sec:related_work}
% What's next 
In the following, we briefly discuss related work from dynamic graph visualizations, the visual analysis of dimensionality reduction methods, and pixel-based visualization techniques.

\subsection{Dynamic Graph Visualization}
% Visual analysis of dynamic graphs  
In many application domains, dynamic graph visualization techniques have recently gained more research attention~\cite{beck2017taxonomy}.
% Why 
Such techniques can be classified into two main categories: animation and timeline visualizations~\cite{beck2017taxonomy}.
% Animations
The animation of large-scale dynamic graphs is often regarded as inadequate due to the cognitive efforts to maintain a mental map~\cite{purchase2006important} and trace changes~\cite{tversky2002animation}.
% Timeline 
On the other hand, timeline visualizations map the graph often to a compact representation to reduce the cognitive efforts and to enable the comparison of periods.
For instance, the parallel edge splatting technique~\cite{burch2011parallel} displays dynamic graphs as a sequence of bipartite graph layouts.
However, even in the improved version, that uses the interleaving concept~\cite{burch2017visualizing}, the identification of temporal patterns remains challenging due to the over-drawing problems between the individual graphs.
% extended massive sequence views
Further, Van den Elzen et al.~\cite{van2013dynamic} proposes to extend massive sequence views and suggests different reordering strategies to minimize block overlaps. 
% Works only for small graphs 
Nevertheless, identifying temporal patterns in dense and large-scale graphs remains challenging due to the overlapping edges, making it difficult to trace changes in the linear-ordering.
% More information 
An extensive survey of further dynamic graph visualization techniques can be found in the surveys of Kerracher et al.~\cite{kerracher2015task}, Beck et al.~\cite{beck2017taxonomy}, and Nobre et al.\cite{nobre2019state}.

% Problem 
In summary, dynamic graph visualizations such as animations and timeline mappings do not scale to long sequences of large-scale graphs due to limited display space~\cite{beck2017taxonomy}.
% Read more here 
% Dimensionality reduction 
Therefore, previous visualization approaches apply dimensionality reduction methods to reduce the complexity of the high-dimensional graph data.

%\subsection{Dimensionality Reduction Based Visualizations}
\subsection{Dimension-Reduced Dynamic Graph Visualization}
% Why 
Visualization approaches based on dimensionality reduction focus on summarizing and abstracting dynamic graphs to highlight temporal and structural changes.
% Dimensionality reduction 
For example, Van den Elzen et al.~\cite{van2016reducing} use dimensionality reduction methods (e.g., t-SNE~\cite{maaten2008visualizing}) to reduce the amount of data and provide an overview of high-level temporal states in a dynamic graph.
% Limitations 
The proposed visual analytics approach, however, depends on the temporal discretization scale and requires feature engineering for embedding the discretized intervals into vectors.
% Time curves 
Time curves~\cite{bach2015time} likewise embed the temporal data in a spatial layout to highlight temporal patterns and anomalies.
% Problems 
Still, time curves heavily depend on feature engineering of the vectors, the quality of the distance metric, and often produce visual clutter for long sequences due to overlapping issues.
% Further reading 
For further reading, we refer to the survey Engel et al.~\cite{engel2012survey}, Sacha et al.~\cite{sacha2016visual}, and the work of Vernier et al.~\cite{vernier2020quantitative}.

% Summary
% Why 
Overall, dimensionality reduction methods reduce the complexity of the dynamic graph data and support the identification of temporal patterns.
% Limitations 
However, such methods frequently fail to capture structural changes as the methods heavily depend on the designed feature vector (e.g., graph metrics).
% Clutter and overlap
Furthermore, such methods do not scale to long sequences due to the visual clutter caused by overlaps in the spatial layout.
% Solution pixel-based visualizations
Pixel-based visualization techniques can be utilized to avoid such over-drawing problems.

\subsection{Pixel-Based Visualization}
% Why 
Pixel-based visualizations effectively use the whole display space and allow us to present large amounts of data without overlap and clutter~\cite{keim2000designing}.
% MotionRugs
For example, Buchmueller et al.~\cite{buchmuller2018motionrugs} highlighted the usefulness of pixel-based visualizations for the visual summarization of changes in spatio-temporal data.
% Gap 
There are, however, only a few pixel-based visualizations for dynamic graphs.
% Example 
For instance, a matrix of pixel-based glyphs can be used to highlight temporal patterns in small social networks~\cite{stein2010pixel}.
% Temporal Treemaps
Furthermore, space-filling temporal treemap visualizations~\cite{kopp2018temporal} can be extended to display medium-sized evolving trees in a pixel-based visualization manner.
% Limitations 
Such temporal treemaps~\cite{kopp2018temporal} and other hierarchy based visual metaphors~\cite{burch2017scalable} are, however, only able to depict evolving hierarchies.
% GraphFlow 
Another pixel-based timeline visualization is GraphFlow~\cite{cui2014let}, which displays graph metrics to highlight structural changes in a dynamic graph.
% Challenges 
The GraphFlow~\cite{cui2014let} method, however, only works for small graphs with a limited number of time steps, and the energy-based visualization also heavily depends on the used graph metric (e.g., node degree) that can fail to capture the overall dynamic phenomena.

\subsection{Delineation to our Work}
% What we do 
In this work, we propose an overlap free multiscale pixel-based visualization that does not require any feature engineering, scales up to long sequences of graphs, and enables us to drill-down into aggregated temporal intervals.
% No feature engineering 
We utilize unsupervised analysis methods (graph embeddings) from the field of machine learning to automatically learn and embed graph structures in low-dimensional space without requiring any features engineering~\cite{zhang2018network}.
% Difference 
Such graph embedding methods stand in contrast to previous analysis methods (e.g., GraphFlow~\cite{cui2014let}) that typically used static or dynamic graph metrics (e.g., diameter~\cite{brandes2004analysis} or change centrality~\cite{federico2012visual}). 
% Highlight the gap 
However, there are currently no visualization techniques that leverage graph embeddings for dynamic graphs even though they have shown to be efficient for various tasks (e.g., in link prediction~\cite{goyal2018graph}).
% Inspiration
Our work was inspired by the stripe-based visualizations of word embeddings that can be used to highlight semantically similar word groups~\cite{shin2018interpreting}.
% Multiscale time
Contrary to previous approaches, \newtechniquename scales to long and large-scale dynamic graphs, providing an encompassing overview of possible temporal aggregation levels.
% Usecase 
In this work, we also investigate how the visual patterns in \newtechniquename can be interpreted.

%% ------
% Multiscale time
% Contrary to previous approaches, we also display the reduced dynamic graph simultaneously at different temporal scales, hence providing an encompassing overview of possible temporal aggregation levels.
% Scaling 
% The multiscale temporal modeling facilitates the pixel-based visualization to scale to long and large-scale dynamic graphs since the precomputed graph embeddings are also small enough to fit into the main memory.
% Usecase 

%% Old stuff 
%% ------
%% Comments 

% \dominik{Title does not state clearly that it is about the visualization of dimension reduced dynamic graphs.}
% \eren{Some suggestions?}
% \dominik{Visualization of dimension-reduced dynamic graphs?}

% \dominik{Why is time curves relevant? It is not primarily related to dynamic graphs and also faces the same issues as any other dimensionality reduction visualization technique. Are there any other approaches that embed dynamic graphs and make them visually accessible?} \eren{They have a section in their paper for networks} \dominik{alright then, just remembered that they applied it to various datasets, not only networks.}

%%%%%%%%%%%%%%%%%%%%%%%%%%%%%%%%%%%%%%%%%%%%%%%%%%%%%%%%%%%%%%%%%%%%%%%%%%%
% Technique Visualization
%%%%%%%%%%%%%%%%%%%%%%%%%%%%%%%%%%%%%%%%%%%%%%%%%%%%%%%%%%%%%%%%%%%%%%%%%%%
\section{Dynamic Graph to Pixel-Based Visualization} \label{sec:main_section}
\begin{figure*}[tb]
 \centering 
 \includegraphics[width=\linewidth]{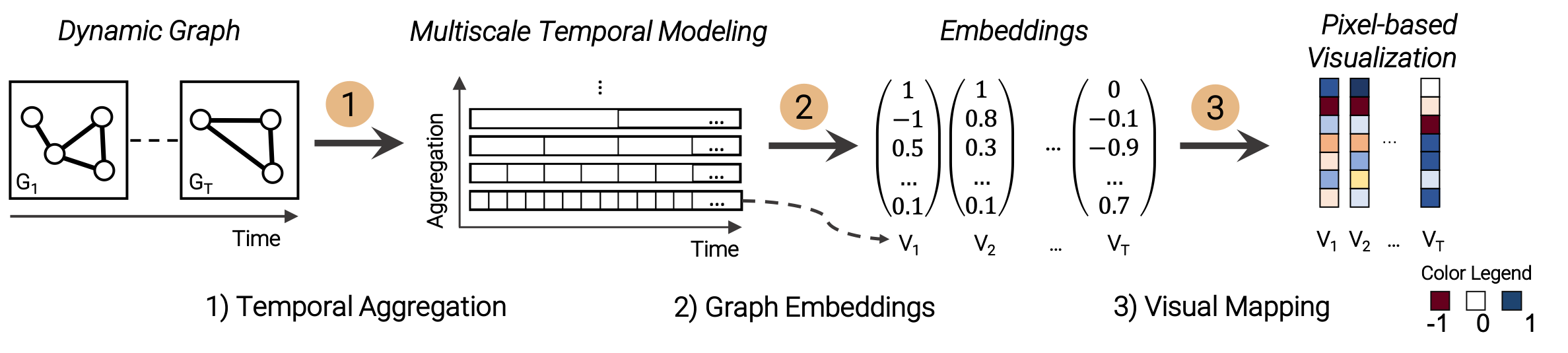}
 \vspace{-0.6cm}
 % \vspace{-1\baselineskip}
 \caption{
 The \newtechniquename enables users to discover similar temporal states. 
 The approach consists of three main adjustable steps (1) temporal aggregations, (2) graph embeddings, and (3) the visual mapping to the pixel-based visualization.
 }
 \vspace{-0.4cm}
 \label{fig:pipeline}
\end{figure*}
% dg2pix
\newtechniquename is a scalable visualization technique to gain an overview of the temporal dimension in long and large-scale dynamic graphs.
% How 
The approach combines temporal aggregations with dimensionality reduction methods (graph embeddings) at multiple temporal scales to reveal temporal patterns, for instance, reoccurring states with similar graph structures. 
% What 
With \newtechniquename, we show how graph embeddings can be interactively used to surpass state-of-the-art visualizations techniques for dynamic graphs by the amount of displayed information.

% The approach 
% Simple steps - but not considered in previous work 
The visualization technique consists of three adjustable steps (see Fig.~\ref{fig:pipeline}).
% Graph embeddings 
The technique's basic idea is to use graph embeddings combined with a pixel-based visualization to present vast amounts of high-dimensional data to support the exploration and summarization of dynamic graphs~\cite{brehmer2013multi}.
% Both approaches 
The (2-3) transformation steps (see Fig.~\ref{fig:pipeline}) are, to the best of our knowledge, not considered in previous literature for dynamic graph visualizations even though graph embeddings outperform many state-of-the-art unsupervised learning methods~\cite{zhang2018network}, and pixel-based visualizations enable to display large amounts of data overlap and clutter-free~\cite{keim2000designing}. 
% What is next 
In the following, we explain the three steps of \newtechniquename and present the implemented prototype.

%% --------
%% --------
\subsection{Multiscale Temporal Modelling}% Why 
Temporal abstraction methods (e.g., temporal aggregations) are applied to dynamic graphs to reduce the amount of data and summarize the changes over time. 
% How 
Typically, such temporal abstraction methods are based on aggregation, such as the supergraph computation, which summarizes intervals by unifying all nodes and edges of a sequence of graphs~\cite{hadlak2013supporting}. 
% Summarization 
Supergraphs provide an overview of temporal intervals by summarizing a sequence of graphs to only one graph with the cost of discarding temporal information~\cite{bender2006art}.
% Discreiziation 
The computation of supergraphs can be seen as a discretization of the temporal dimension.
% Usefulness 
However, the usefulness of such a temporal discretization depends on many aspects, for instance, graph size, frequency of topological changes, and the temporal aggregation scale~\cite{bender2006art}.
% Example 
For example, a fine-grained temporal aggregation into supergraphs results in various intervals with little information unable to provide an overview. 
In contrast, coarse-scale aggregation produces only a few supergraphs, which may contain a high variance, where important intervals may remain hidden.
% Finding the right level of abstraction 
Moreover, finding the optimal fixed interval length for analysis depends on the user task at hand~\cite{devineni2017one}. 

% Consequence 
We compute supergraphs at different temporal aggregation scales to enable users to explore temporal states at multiple temporal granularities interactively.
% Why 
In contrast to previous discretizations of time that use uniform or non-uniform temporal granularities (time-slices)~\cite{wang2019nonuniform}, we propose to recursively partition the data using uniform time-slicing methods and compute for each interval a supergraph. 
% Example 
For example, the recursive supergraph generation can be done based on the cyclic division of the time, such as the division into a year, months, and days. 
% What 
For domains with no reasonable temporal partitioning, we propose the following default dynamic graph coarsening approach.
% How 
The default approach slices $T$ time steps of the temporal dimension recursively into intervals of length $2^l$ with $l$ being the level $l \in 1,..., \lceil log(T) \rceil$.
% Result
The resulting levels contain at the lowest level one intervals of length one and the highest level $\lceil log(T) \rceil$ a supergraph of all graphs. 
% Supergraphs 
We compute a supergraph for each of these intervals, which results in $\lceil T / 2^{l-1} \rceil$ supergraphs for each the level $l$.
% Hierarchy 
The multiscale temporal modeling computes $\lceil log(T) \rceil$ levels of granularity having overall $(2 \cdot T) + 1$ supergraphs.

% Why 
In summary, the temporal multiscale modeling essentially recursively coarsens the dynamic graph into supergraphs, which are used in combination with the original evolving graphs in the next step to learn the similarities between graphs in a latent space.
% How 
Our multiscale temporal modeling was inspired by Elmqvist and Fekete~\cite{elmqvist2010hierarchical} hierarchical aggregation, which enables us to turn visualizations into multiscale (multiresolution) approaches that scale better to large datasets.
% Drill down and roll up 
The multiscale temporal modeling is later used to perform unbalanced drill-down and roll-up operations.

%% --------
%% --------
\subsection{Graph Embedding}
% Next step - why 
In the second step, we apply dimensionality reduction methods (graph embeddings) to all generated snapshots to learn the similarity between graphs and reduce the high-dimensional data to low-dimensional vectors. 
% Why 
We apply graph embeddings (e.g., graph2vec~\cite{narayanan2017graph2vec}) as they are scalable to large-scale dynamic graphs, outperform state-of-the-art methods in the field of unsupervised learning, do not require feature engineering, and are small enough to fit into main memory for interactive visual analysis~\cite{zhang2018network}.
% Example 
For example, graph2vec outperforms graph kernels, and substructure embedding approaches for classification tasks on large graph datasets~\cite{narayanan2017graph2vec}. 
% What 
A graph embedding can be seen as a function $f: V \rightarrow \mathbb{R}^d$ that maps a set of vertices (e.g., random walks) to a $d$ dimensional embedding. 
% Latent space 
Typically, the embeddings of the latent space $\mathbb{R}^d$ are used to gain insight into the data and for further standard machine learning tasks.   
For example, we cluster the embeddings to visualize and gain an overview of similar temporal states.
% Advantage 
In contrast to previous dynamic graphs visualization approaches (e.g., van Elzen et al.~\cite{van2016reducing}), which depend strongly on the used graph metric, unsupervised graph embeddings do not require any feature engineering, are task agnostic and data-driven.
% Reason 
An advantage of graph embeddings is that the methods learn similarity between graphs in the latent space by approximating different graph metrics~\cite{bonner2019exploring}. 
% Also, why graph embeddings and not vertex embeddings?
Furthermore, we employ graph embeddings instead of node embeddings because graph embeddings only compute one vector for a given time step and therefore scale to large datasets.

% How 
We utilize the three recently proposed graph embeddings~\cite{karateclub} for our approach: graph2vec~\cite{narayanan2017graph2vec}, GL2Vec~\cite{chen2019gl2vec}, and FGSD~\cite{verma2017hunt} as the approaches have moderate run-time complexities.
% Supergraph embeddings
We compute embeddings of all $2T+1$ supergraphs of the multiscale temporal modeling step and embed the graphs, as suggested by Bonner et al.~\cite{bonner2019exploring}, into the range of $50-300$ dimensions.
% Normalization 
Per default, we embed each graph to a vector of 128 features and apply $L_2$ normalization to the embeddings. 
% Why 
The normalization maps the vectors to unit length and enables us to use cosine similarity instead of the dot product as a distance measurement~\cite{levy2015improving}. 
% Vectors 
The vectors are later displayed in the pixel-based visualization as pixel-bars to identify changes in dynamic graphs visually. 
% Example
In Fig.~\ref{fig:use-case-synthetic}, three different graph embeddings of a synthetic dataset are presented, highlighting that the proposed methods capture temporal states in dynamic graphs.
% Summary 
In our discussion (see Sec.~\ref{sec:discussion}), we elaborate on the input parameters and the scalability of such graph embeddings.

%% --------
\subsection{Pixel-based Visualization}
% Pixel-based visualization 
In the last step, we visually encode the embeddings into dense pixel-based visualizations to provide an overview of the temporal dimension and reveal similar graphs.
% Why embeddings 
We particularly visualize the embeddings as they are compact encodings of the structural information of each graph. 
% How 
The embeddings are displayed as linearized pixel-bars that are basically grid-based columns in which each rectangle (pixel) is a feature of the embedded vector.
% Coloring 
The technique colors each pixel by the feature's value by using a diverging color scheme from ColorBrewer~\cite{harrower2003colorbrewer}.
We utilize a global segmented color scheme with two distinct values to support the comparison task~\cite{tominski2008task}.
% Temporal dimension 
We sequentially order each displayed pixel-bar (graph embedding) by time, creating a dense pixel-based visualization.

 % Ordering of the grid-based view
The y-axis ordering of the colored pixel-bars is per default based on the linear order of the vector. 
% The problem 
The challenge of finding a useful linear order to highlight particular patterns visually can be mapped to the linearization problem~\cite{behrisch2016matrix}. 
 % Ordering 
We are utilizing different reordering algorithms to improve the global ordering of the embeddings to emphasize different patterns along both axes of the pixel-based visualization. 
% Example
For example, we apply clustering algorithms to all displayed data features to group and arrange similar features over time.
We discuss different reordering strategies in Sec.~\ref{sec:reordering}. 

% Multiscale 
We utilize the computed supergraphs of the multiscale temporal modeling to present the data at multiple user-defined levels of temporal aggregation (see Fig.~\ref{fig:zoom-level}). 
Such a multiscale (multiresolution) visualization helps to set detailed abstraction levels into the overall temporal context~\cite{elmqvist2010hierarchical}.
% Example 
For example, the visualization technique presents 1000 supergraphs as pixel-bars instead of several thousand individual graphs and enables users to drill-down into intervals.
% How 
We limit the number of depicted grid-based columns to the available horizontal pixels of the screen space to address our approach's visual scalability, which means that the minimum width of a pixel-bar is precisely one pixel. 
% Display space  
If a user drills down and reaches the limit of screen space pixel, he has to coarsen temporal intervals to reduce the number of overall displayed pixel-bars.
% Next section 
Next, we describe our implemented prototype and available interaction methods.

%% --------
%% --------
\subsection{Prototype}
% Why 
The \newtechniquename prototype implementation~\footnote{\href{https://github.com/eren-ck/dg2pix}{https://github.com/eren-ck/dg2pix}} enables us to explore the temporal changes of large scale dynamic graphs.
% What
In the following, we briefly introduce the two main linked views of the prototype.

\begin{figure}[tb]
 \centering 
 \includegraphics[width=0.96\linewidth]{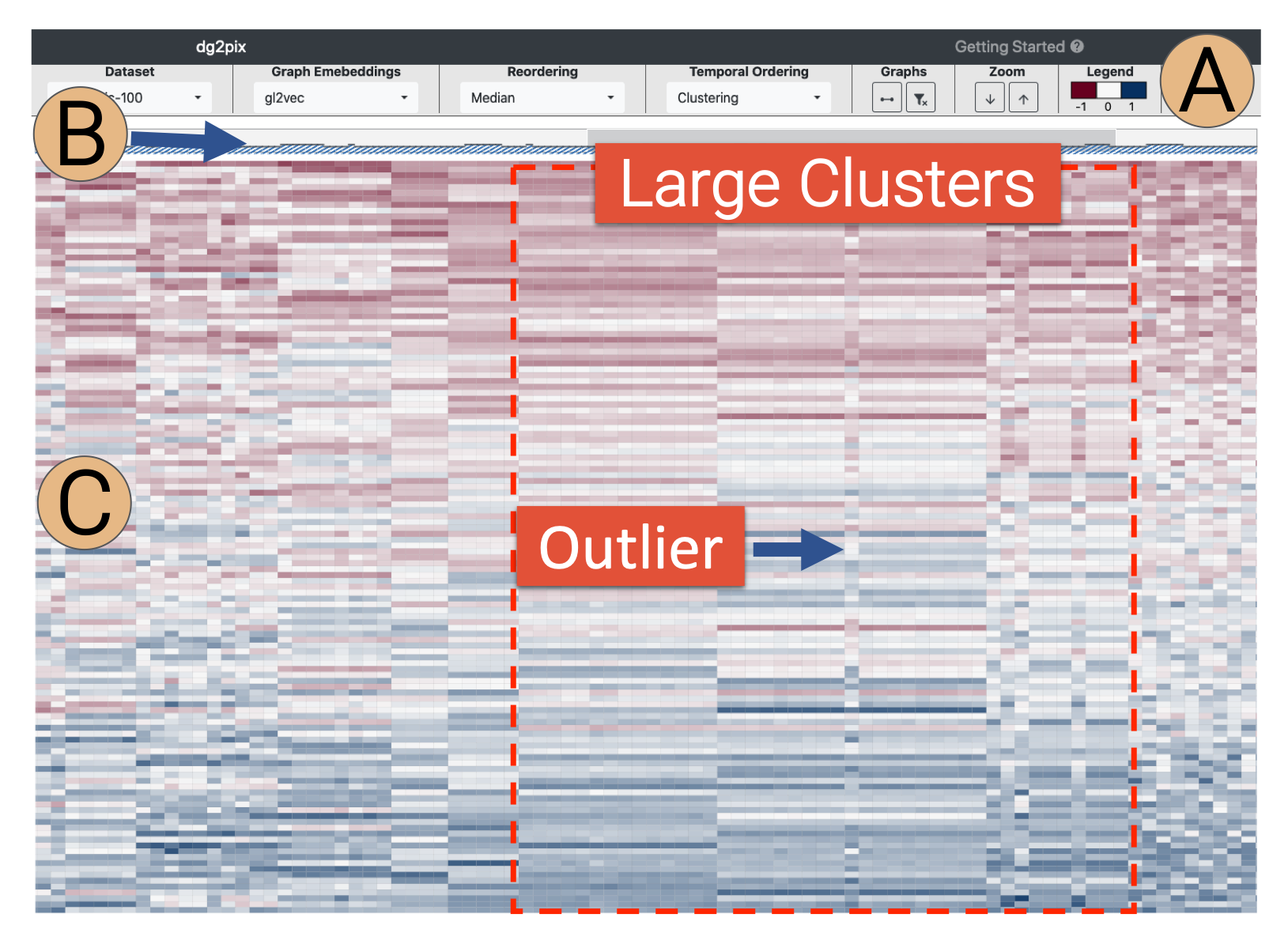}
 \vspace{-0.2cm}
 \caption{
 The \newtechniquename component consists of three views a toolbar (A), the zoom context bar (B), and the pixel-based visualization (C).
 % What
 In (C) the embeddings (GL2Vec~\cite{chen2019gl2vec}) of a synthetic dynamic graph (SBM\cite{holland1983stochastic}) with reoccurring states are depicted. 
 The displayed embeddings of the x-axis are clustered (HDBSCAN~\cite{campello2013density}), and y-axis ordering is based on the median of each vector attribute.
 % Visible 
 The reordering and clustering of the synthetically created reoccurring states highlight large clusters of similar graphs and outliers in the temporal data.
 }
 \vspace{-0.4cm}
 \label{fig:prototype}
\end{figure}

\begin{figure}[tb]
 \centering 
 \includegraphics[width=\linewidth]{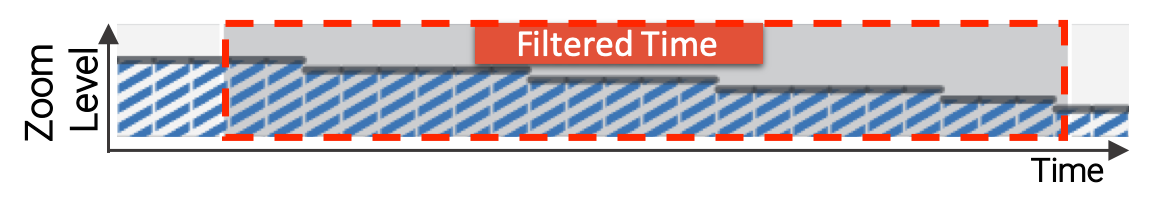}
 \vspace{-0.8cm}
 \caption{
 % Why 
 The zoom context bar enables us to investigate the zoom level for an individual and multiple pixel bars. 
 Further, it allows filtering time intervals for vertical and horizontal navigation. 
}
 \vspace{-0.6cm}
 \label{fig:zoom-level}
\end{figure}

\looseness=-1
The \textbf{\newtechniquename view} (see Fig.~\ref{fig:prototype}) consists of a \textit{toolbar} (A), a \textit{zoom context bar} (B), and the \textit{pixel-based visualization} (C).
% Why toolbar 
The \textit{toolbar} allows selecting and presenting various graph embeddings for particular datasets, including choosing different training epochs and applying automated analysis methods.
For example, the x-axis can be reordered based on a clustering of the graph embeddings (see Sec.~\ref{sec:reordering}).  
% More 
Furthermore, the toolbar enables us to change the temporal granularity of intervals (drill-down and roll-up) and display selected graph embeddings in the graph view.
% Zoom context bar 
The \textit{zoom context bar} presents additional information for the vertical and horizontal temporal navigation and provides an overview of the displayed temporal interval and granularities.
% How 
The view (see also Fig.~\ref{fig:zoom-level}) displays for each pixel-bar the corresponding temporal granularity as a zoom bar (rectangle).
The height is mapped to the zoom level, meaning the zoom bars of low levels of temporal granularity are rather small.
% Mouse over 
The zoom bars are always ordered by time and enable us to relate the potentially reordered pixel-bars to their overall temporal context via brushing and linking.
% Period 
The zoom context bar also allows for selecting and filtering periods of the pixel-based visualization, allowing navigating horizontally along the temporal dimension.

% Pixel-based visualization 
The \textit{pixel-based visualization} displays per default the medium zoom level of the graph embeddings ordered by time. 
% How 
The view allows us to select individual and multiple pixel-bars and adapt the temporal granularity by drilling-down a lower temporal granularity or coarsening the temporal dimension (roll-up).
% Linkage to zoom context bar 
The view is also directly linked to the zoom context bar, enabling us to keep an overview and relate the pixel-bars to the temporal dimension.
% Reordering strategies
The x- and y-axis of the pixel-based visualization can also be reordered using different reordering strategies to highlight clusters and similarities between the embeddings (see Sec.~\ref{sec:reordering}). 
% Graph view transition 
Furthermore, multiple pixel-bars can be selected to display the underlying supergraphs and graphs in the second main view.

% Why 
The \textbf{graph view} allows us to display the underlying graph data of the selected pixel-bars as a supergraph to highlight and compare the intersections and disjoint nodes and edges between the graphs in the temporal data.
% Coloring 
The supergraph nodes and edges are colored using two graph set operations on all selected time steps to highlight similarities and differences.
% Intersection 
The applied set operations are intersection (orange) and disjoint (blue) on all nodes and edges of the selected time steps. 
% The main goal 
The main goal of this comparison using set operations is to investigate the changes in the temporal graph, which helps to identify and interpret which graph structures were preserved in the latent space.
% Layout 
The view uses per default for all time steps one precomputed graph layout (Fruchterman-Reingold~\cite{fruchterman1991graph}) by computing a supergraph for the whole dynamic graph.
% Advantage 
We facilitate one global layout for the whole dynamic graph to preserve the user's mental map~\cite{beck2009towards}.
% Semantic zooming 
The graph view can also be explored via semantic zooming to explore particular graph structures, such as specific node and link attributes. 

\section{Visual Interpretation of dg2pix} \label{sec:interpretation}
% why 
\newtechniquename provides a scalable overview of the temporal dimension to emphasize underlying changes in dynamic graph data.
% How 
The main idea of the approach is to learn and display low-dimensional embeddings of graphs that capture the similarity between graphs in a latent space. 
% Problem 
However, the interpretation of such embeddings in the latent space remains challenging as the meaning of particular numeric values cannot be directly mapped to topological features of the graph.
% Main problems 
For example, the specific meaning of a dimension value of $0.3$ of an embedding with 128 dimensions remains unanswered. 
% Conclusion 
Consequently, the abstractness of what low and high values of each dimension encode poses a challenge to understand and map the patterns in \newtechniquename to topological changes in the evolving graph.
% Solution
In previous work, typically, 2D visualizations are used to interpret and understand such latent space~\cite{liu2019latent}.
% Example  
For instance, the Embedding Projector~\cite{smilkov2016embedding} by Google Brain uses projections (e.g., t-SNE~\cite{maaten2008visualizing}) to present word embeddings as 2D and 3D scatterplots.
% Problem 
However, such simple 2D visualizations discard latent space information as the $d$-dimensional embeddings are again reduced into a 2D embeddings for the visual representation.
% Next 
The following section describes the underlying challenges of visualizing latent spaces, the interpretation of visual patterns, and different reordering strategies to highlight temporal changes.

%% --------
\subsection{Latent Space Visualizations}
% Interpreting latent spaces 
Recently, the visual analysis of latent spaces (embedding spaces) has gained research interest~\cite{liu2019latent}.
% Multidimensional data visualization 
For example, ad-hoc dimensionality reduction methods (e.g., PCA, t-SNE~\cite{maaten2008visualizing}, or UMAP~\cite{mcinnes2018umap}) are often applied to display neighbors in the latent space in 2D space.
% The goal 
The latent space representation central goal is to provide more insight into the underlying embedded data and enable the qualitative interpretation of the learned embeddings~\cite{liu2019latent}.
% Example 
Heimler and Gleicher~\cite{heimerl2018interactive}, for instance, describe tasks for word embeddings and display words in a matrix-based view to highlight co-occurrences between words.
% Extension 
Further, Liu et al.~\cite{liu2019latent} describe a set of tasks for exploring latent spaces and present a cartography system to visually investigate relationships between data points and compare attributes of vectors (e.g., word embeddings).
% Graph embeddings  
The visual analysis of latent spaces currently remains the primary method to investigate and interpret graph embeddings.
There has been little theoretical work to prove that such embeddings approximate and learn different graph metrics~\cite{bonner2019exploring}.
% Example 
For example, EmbeddingVis~\cite{li2018embeddingvis} enables the comparison of different latent spaces of node embeddings to investigate which node metrics are preserved by applying regression. 

% Pixel-based visualizations 
In contrast to all previous approaches, our primary goal is to generate a visual summary of the temporal dimension that helps to understand and highlight temporal states in the evolving data. 
% Pixel-based visualization 
We display the embeddings with all their dimensions to visually compare similarities and apply reordering strategies to present changes in the latent space.  
% Interpretation
Our approach also allows us to present the underlying graphs in combination using graph set operations (e.g., union or intersection) to help interpret and compare the latent space with the original evolving graph data.
% Next
Next, we elaborate on how \newtechniquename can be interpreted, and automatic approaches can be used to find similar temporal states.
% ------
\subsection{Interpretation of Visual Patterns}
% Why 
Graph embeddings are machine learning models that produce abstract low-dimensional vector representations for graphs that are difficult to interpret, as the individual values of the dimension themself have no exact interpretation~\cite{salehi2017properties}.

% Reasons 
\textbf{Challenges} The reasons for interpretation challenges arise from the stochastic algorithms (e.g., graph2vec~\cite{narayanan2017graph2vec}), which utilize non-transparent neural networks with hyperparameters~\cite{salehi2017properties}. 
% Unitary rotations 
Further, the embeddings can be changed with unitary rotation, which completely transforms each dimension's values while preserving the latent spaces distances.
% The challenge 
Therefore, the complexity of interpreting graph embedding dimensions can be compared to the efforts to understand activations in neural networks for image classification~\cite{salehi2017properties}.
% More 
Nevertheless, recent experiments~\cite{salehi2017properties, bonner2019exploring} indicate that graph embedding methods learn to approximate various topological features of graphs. 
% Effectiveness 
Therefore, we utilize and visualize graph embeddings to highlight changes in dynamic graphs as the methods have shown to be effective feature spaces for various graph mining tasks, such as classification of graphs~\cite{verma2017hunt, tsitsulin2018netlsd, chen2019gl2vec}.

% Identification of patterns 
\textbf{Interpretation}
% Visualization 
The pixel-based visualization enables us to perceive similarities and differences between embeddings to provide an overview of the dynamic graph. 
% Why 
Generally, the visualization of embeddings can reveal relationships in the latent space, as shown by Shin et al.~\cite{shin2018interpreting} for the comparison of semantically similar word embeddings.
% How 
The graph embeddings can only be interpreted in relation to other embeddings by investigating the pairwise similarity between embeddings.
% Interpretation 
More specifically, if two subsequent graph embeddings in the dynamic graph are, to some extent, similar to each other, then the original graphs are also similar to one another. 
Also, vice versa, if two successive embeddings are different, then the two embedded graphs are dissimilar to some extent.
% Detect change 
Therefore, we can use the embeddings to examine and highlight changes and temporal states in a dynamic graph even though we cannot interpret the individual values of particular dimensions.

% Comparing embeddings 
\textbf{Visual Comparison of Embeddings}
Consequently, the human-centric visual analysis of temporal states (e.g., reoccurring graphs) can be mapped to distinguishing similar pixel-bars in the \newtechniquename.
% Example 
For instance, Fig.~\ref{fig:prototype} displays a large block of similar pixel-bars with an apparent outlier in-between.
% User
The visual analysis of pairwise similarities between pixel-bars enables identifying temporal changes and states in the underlying dynamic graph.
% Cognitive efforts 
However, the cognitive efforts to compare multiple pixel-bars are high since the user has to simultaneously relate numerous dimensions of different embeddings.
% Automatic solution 
The pairwise similarities between multiple embeddings can also be computed using the cosine similarity.
% Solution
We, therefore, propose to use automatic methods to sort and cluster similar rows and columns in the pixel-based visualization to enable the identification of temporal states (e.g., outliers) in the dynamic graph.

% Graph View Linkage 
\textbf{Explainability}
We also compare the underlying graph structures of embeddings in the graph view against each other, intending to generate new insight into the latent space.
% Main goal 
For instance, displaying the graph data helps to explain potential reasons and impacts of graph features on particular values for individual dimensions.
% Example 
Overall, both the pixel-based visualization and the graph view can help to understand and explain the semantic meaning of high and low values of particular dimensions to gain new insight into graph embeddings, which are currently black-box models \cite{salehi2017properties}.

\begin{figure*}[tb]
 \centering 
 \includegraphics[width=\linewidth]{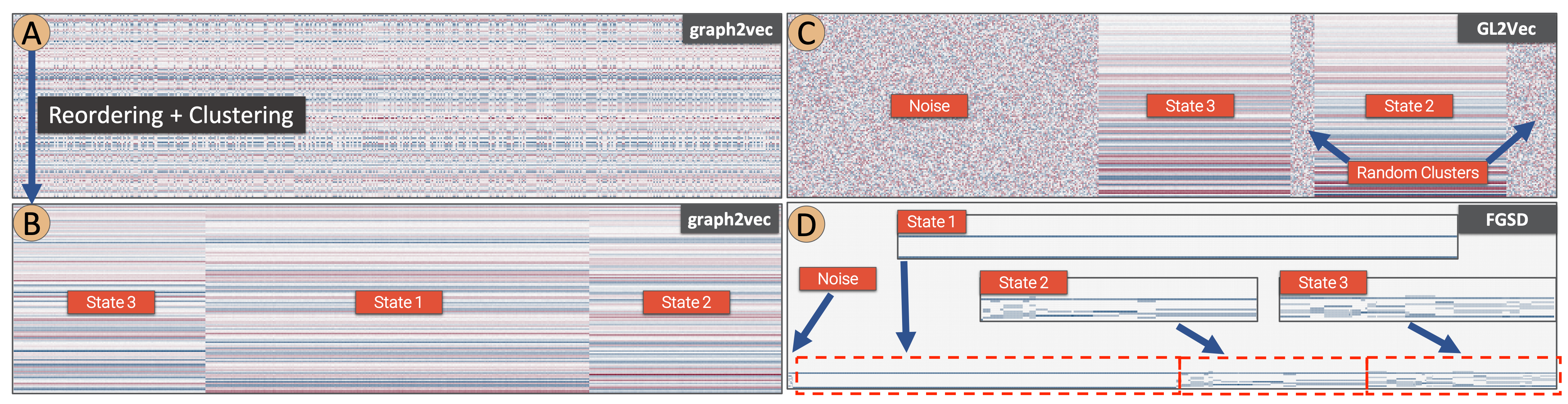}
 \vspace{-0.4cm}
 \caption{
    % Why 
    The synthetic dynamic graph described in the use cases (see Sec.~\ref{sec:use-case-1}) displays three different graph embeddings with a ground truth of three temporal states.
    % How 
    (A) presents the dynamic graph using graph2vec~\cite{narayanan2017graph2vec}, and (B) shows the same data with the three temporal states. 
    % What
    In (B-D), the same reordering strategies were applied to highlight the temporal states. 
    % C and D
    The two other graph embeddings, (C) GL2Vec~\cite{chen2019gl2vec} and (D) FGSD~\cite{verma2017hunt}, are partially able to learn and highlight the three temporal states in the synthetic dynamic graph.
 }
 \vspace{-0.2cm}
 \label{fig:use-case-synthetic}
\end{figure*}
% ------
\subsection{Reordering Strategies} \label{sec:reordering}
\newtechniquename was designed to scale to large-scale dynamic graphs and provide a visual summary of the temporal data. 
% Challenge 
However, temporal states can remain hidden and difficult to identify due to the sheer amount of visualized data, for instance, if single reoccurring pixel columns correlate with other prominent states. 
% Reordering 
Applying different reordering strategies to the embedding can reveal such otherwise hidden temporal states. 
% Example 
For example, clustering reordering the displayed pixel columns (x-axis) will highlight similar graph structures. 
% Consequence
Therefore, we provide users with the option to apply reordering strategies to reveal similar patterns along both axes.

% Methods 
We provide global reordering strategies for the dimensions of the embeddings (y-axis) and the temporal dimension (x-axis).
% Reordering 
In general, identifying an optimal ordering for our pixel-based visualization is known to be NP-Hard since the issue can be mapped to the problem of reordering (linearization) of rows and columns in matrices~\cite{behrisch2016matrix}.
% Main problems 
For the reordering of matrices, various reordering strategies (layouts) have been proposed to highlight different patterns (e.g., block patterns~\cite{behrisch2016matrix}).  
% X-axis 
We provide for the reordering of the embedding dimensions (y-axis) several heuristics based on a statistical metric of each row. 
% First y-axis solution 
For example, before the $L_2$ normalization, the y-axis can be sorted by the median value for each dimension of the displayed embeddings to highlight block and band patterns~\cite{behrisch2016matrix}.
% Second y-axis solution 
Furthermore, the prototype allows us to reorder the dimensions (rows) of the pixel-based visualization using the mean, minimum, maximum, variance, and standard deviation of the depicted rows.

% Y-axis 
We also provide two reordering strategies for the temporal dimension (x-axis) to identify similar temporal states by computing clusters and reordering based on the distances to one particular column (similarity search). 
% Clustering approach 
The clustering uses HDBSCAN~\cite{campello2013density} for the displayed embeddings facilitating the cosine-similarity as a distance measurement.
% Why HDBSCAN 
We employ HDBSCAN~\cite{campello2013density} as the approach aims to find the result with the best stability over different epsilons parameters and accordingly detect clusters with varying densities.
% The ordering 
The clustering results are displayed by grouping and highlighting the pixel-bars according to their clusters using a grey bounding box. 
% Clusters 
For instance, the clusters are reordered using the median time of all embeddings, and the underlying embeddings of a cluster are again sorted by time. 
% Second 
Second, we enable to reorder the y-axis based on the distance to a particular embedding. 
The resorting places an embedding to the first position and afterward ranks the presented embeddings by the distance to the selected embedding (similarity search). 
% Why useful 
This reordering enables us to compare one particular embedding in time with all other graph embeddings in detail.

% What is the outcome 
Overall, using such reordering strategies for both axes can help users understand how the ordering influences the visual patterns, can group, and rank similar temporal states to explore the latent space in more detail.

%% ------
%% Old stuff 

%% ------
%% Comments 
% \dominik{I think you mentioned all strategies to interpret patterns. However, because you claim this as a contribution, I think it is missing a bit of methodology. lets have a quick call in that regard.}

%%%%%%%%%%%%%%%%%%%%%%%%%%%%%%%%%%%%%%%%%%%%%%%%%%%%%%%%%%%%%%%%%%%%%%%%%%%
% Use Case
%%%%%%%%%%%%%%%%%%%%%%%%%%%%%%%%%%%%%%%%%%%%%%%%%%%%%%%%%%%%%%%%%%%%%%%%%%%
\section{Use Cases}\label{sec:evaluation}
% Why 
In the following section, we apply \newtechniquename to synthetic and a real-world dynamic graph to demonstrate how the approach can be used to gain an overview and provide insight of the temporal changes and reoccurring states in evolving graphs.

%%% ----
\subsection{Synthetic Dynamic Graphs}
\label{sec:use-case-1}
% Why 
We generated synthetic dynamic graphs, with known ground truths, to show the applicability and the usefulness of \newtechniquename. 
% How 
For example, we created different datasets with the Stochastic Block Model (SBM)~\cite{holland1983stochastic} with a fixed amount of nodes for each time step, a varying number of edges, and multiple temporal states (see Fig.~\ref{fig:teaser}). 
% What is next 
We elaborate on the results of one dynamic graph to show how the approach can be used to identify states in large-scale graphs. 

% The data 
The synthetic dynamic graph consists of 1000 time steps, 1000 nodes, more than 30 million edges, and three reoccurring temporal states. 
% What exactly 
We facilitated the SBM to create three states with different numbers of clusters (blocks), a slightly varying number of nodes (up to 50) per cluster, and minor edge density changes (internal and external).
% The dataset 
The dynamic graph consists of randomly shuffled data of 500-time steps with two clusters of nodes, 250-time steps with three clusters, and 250-time steps with four clusters.
% Randomly shuffled 
The dynamic graph was embedded with three different graph embeddings with the following parameters: 
\begin{itemize}
    \setlength\itemsep{0cm}
    \item \textbf{graph2vec}~\cite{narayanan2017graph2vec}: 1000 epochs, 0.02 learning rate, 2 Weisfeiler-Lehman iterations, and 128 dimensions.
    \item \textbf{GL2Vec}~\cite{chen2019gl2vec}: 1000 epochs, 0.02 learning rate, and 128 dimensions.
    \item \textbf{FGSD}~\cite{verma2017hunt}: 128 number of histogram bins with a the histogram range of 20.
\end{itemize}
% Figure
The Fig.~\ref{fig:use-case-synthetic} (A-D) shows the resulting \newtechniquename of the synthetic dynamic graph.
% A-B
In Fig.~\ref{fig:use-case-synthetic} (A), the randomly shuffled data is displayed using the graph2vec~\cite{narayanan2017graph2vec} embeddings, and in (B) the same pixel-bars are presented after the application of reordering strategies. 
% Why useful 
We reordered the embeddings (x-axis) based on the clustering of the embeddings (HDBSCAN~\cite{campello2013density}), and the rows were globally sorted based on the standard deviation of each row (ascending). 
% The reordering 
The reordering strategies help to identify temporal and reoccurring states (e.g., clusters) by grouping similar and dissimilar pixel-bars and their respective rows together. 
% Example
For example, sorting the rows by the standard deviation of each row allows users to compare and identify the embedding dimensions that primarily distinguish temporal states.
% Figure B 
In Fig.~\ref{fig:use-case-synthetic} (B), the three temporal states are visible, which can be verified by displaying the underlying graph structures in the graph view. 
% Advantage 
Accordingly, graph2vec has managed to learn the temporal states encoded in the underlying ground truth.

% GL2vec
In contrast, GL2Vec~\cite{chen2019gl2vec} was not able to distinguish the three temporal states (see Fig.~\ref{fig:use-case-synthetic} (C)). 
The same reordering strategies result in only two visible temporal states. 
% Result
The GL2Vec model learned to distinguish the states with the three and four clusters, however, the model was not able to distinguish the larger group of two clusters (500-time steps) in the latent space.
% What it did 
The clustering grouped the first temporal state as the visible block of noise and identified two similar states in the ground truth as two different clusters.
The GL2Vec model potentially requires a different learning rate or more epochs to distinguish the third state in the latent space.
% Figure D 

In Fig.~\ref{fig:use-case-synthetic} (D), the FGSD~\cite{verma2017hunt} is displayed which approximately learns the three temporal states. 
% States 
Compared to the first two methods, the FGSD model embeds the dimensions only to a positive range (blue color), and only seven dimensions of the embeddings contain values.
% The visible 
The method is almost able to distinguish all three clusters except for a little bit of noise, which can be verified by visualizing the graphs in the graph view. 
% Contrast 
In contrast to the other graph embeddings, the FGSD model results in considerable white space that can be removed by deleting rows that do not contain any values.

\begin{figure}[tb]
 \centering 
 \includegraphics[width=\linewidth]{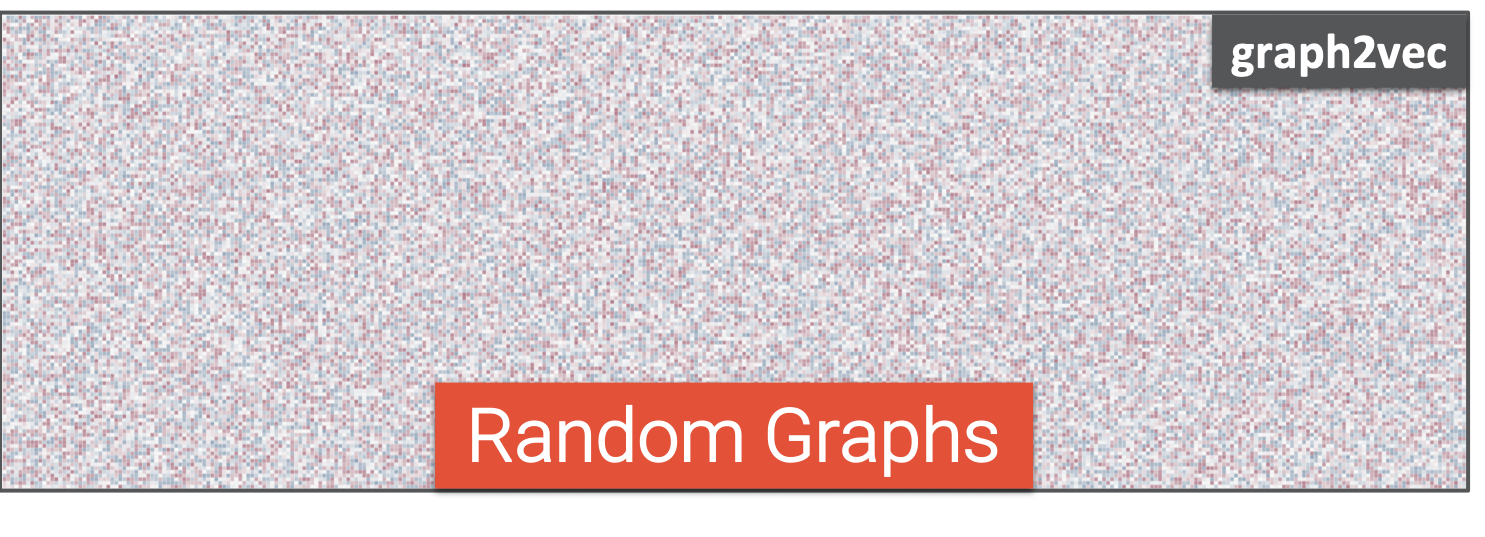}
 \vspace{-0.6cm}
 \caption{
    % Why 
    The synthetic random dynamic graph contains no known ground truth, and the proposed approach produces no visible patterns.
 }
 \vspace{-0.4cm}
 \label{fig:use-case-random}
\end{figure}

% Random graphs
In addition to the different synthetic graphs with known ground truth, we also created random dynamic graphs with different graph generators to confirm that the visible patterns are not arbitrarily learned in the latent space during the training process. 
% How 
For instance, Fig.~\ref{fig:use-case-random} shows a dynamic graph with 1000 randomly generated connected Watts–Strogatz small-world graph~\cite{watts1998collective} with 2000  nodes (between 5-50 nearest neighbors), and $<0.1$ edge probability edges for each time step. 
The same reordering strategies, as in Fig.~\ref{fig:use-case-synthetic}, were applied, and the resulting \newtechniquename shows the graph2vec (1000 epochs) embedding, which does not contain any visible patterns as the model was not able to learn the similarities between the random graphs in latent space.

%%% ----
\subsection{Evolving Social Network}
\label{sec:use-case-2}
% Why
Next, \newtechniquename is applied to a real-world, large-scale social networks.
% How 
We describe the temporal visual analysis of the website Reddit~\cite{kumar2018community} to discover structural and temporal changes as well as reoccurring states between social network communities (subreddits) during the 2016 US presidential elections.
% Elections 
In the following, we describe the analyzed dataset, highlight the main task and challenges for the analysis of such data, and how \newtechniquename can be used to provide an overview of the temporal changes.

% Dataset 
\textbf{Reddit Data}
% Reddit dataset 
Reddit is a social news aggregation website with approximately 440 million users as of 2020. 
The website is made up of subreddits in which users post content (e.g., images or links to news sites) and upvote posts based on a voting based system to rank interesting content for each subreddit.
% What 
The dataset~\cite{kumar2018community} is a dynamic hyperlink graph and consists of subreddits (nodes), and time-stamped hyperlinks (edges). 
% Dataset 
The analyzed data contains hyperlink graphs grouped by hours from the 1st January 2016 to 30th November 2016 in which the election campaign for the 2016 presidential election took place. 
% The data 
The dynamic graph consists of 7974 graphs, 18546 nodes (subreddits), and 88328 edges (hyperlinks) between the subreddits with either positive or negative sentiment. 
% Graph embeddings 
We computed the following three graph embeddings graph2vec~\cite{narayanan2017graph2vec}, GL2Vec~\cite{chen2019gl2vec}, and FGSD~\cite{verma2017hunt} with the same input parameters as described in Sec.~\ref{sec:use-case-1}.
% Validation 
We verified the resulting insight by comparing the identified changes and states of the underlying evolving hyperlink graphs to the real historic news coverage of the presidential elections.

% Tasks 
\textbf{Tasks and Challenges}
% Why
The visual analysis of a social network data aims to provide an overview of structural changes over time, temporal states (e.g., reoccurring graph structures), and outlier graphs in the evolving data (e.g., political scandals).
% Challenge
However, gaining an overview of large-scale social media data is challenging as it requires to visualize structural as well as temporal changes simultaneously and to identify suitable temporal analysis scales for changes and states of varying temporal length.
% More problems 
Furthermore, the size and complexity of social networks pose another challenge to visualize the evolving data since there is a trade-off between the visualization of the detailed graph structure for each time step and presenting the overall evolving graph properties.
For instance, animations display each graph of the data in detail, however, animations are considered to be unsuited to provide an overview of long periods due to cognitive efforts to keep track of changes~\cite{tversky2002animation}.
% Our method 
In contrast to previous approaches, we model and embed dynamic graphs at multiple temporal scales to enable the multiscale temporal analysis of long as well as large-scale dynamic graphs.   

\begin{figure*}[tb]
 \centering 
 \includegraphics[width=0.98\linewidth]{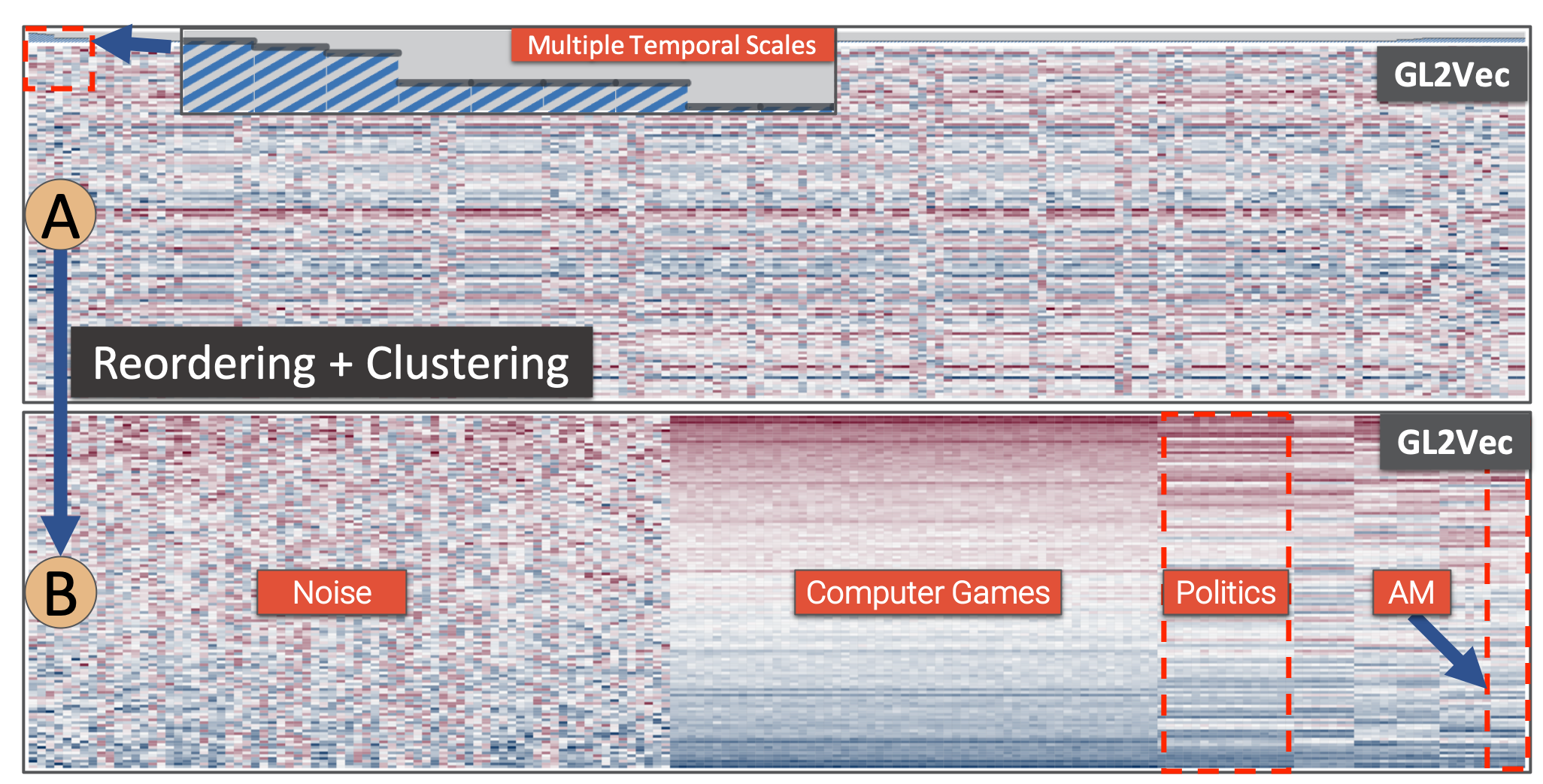}
 \vspace{-0.4cm}
 \caption{
    The Reddit data described in the use cases (see Sec.~\ref{sec:use-case-2}) presents the election week of the 2016 US presidential election using GL2Vec~\cite{chen2019gl2vec} embeddings.
    % How 
    The multiscale temporal modeling was used to display dill-into the election week and aggregate other intervals into supergraphs. 
    % How 
    (A) displays the evolving social networks sorted by time, and (B) shows the same data after applying reordering strategies to emphasize temporal states. 
    % What
    We linked the clusters of embeddings to hyperlinks between different communities of subreddits, for example, computer games related topics, political topologies, or morning graphs structures (AM). 
 }
 \vspace{-0.5cm}
 \label{fig:use-case-reddit}
\end{figure*}

% Tasks 
\textbf{2016 US Presidential Election}
% Why 
We begin by investigating the week before and the week during the 2016 US presidential elections (8th November 2016) to identify graphs with political subreddits in the temporal data. 
% Initial setup 
Per default, the prototype displays 400 pixel-bars of the middle level of temporal granularity using the graph2vec~\cite{narayanan2017graph2vec} embeddings.
% Zooming 
First, we use the multiscale temporal modeling to concentrate on the election weeks in November 2016. 
% Zooming 
We aggregate the pixel-bars before October into aggregated supergraphs (roll-up), and further split (drill-down) the election weeks into the lowest temporal granularity of one hour.
% Graph embeddings 
We display different graph embeddings to examine the resulting pixel-bars during the election week period visually. 
% Decision 
We decide to use the GL2Vec~\cite{chen2019gl2vec} embeddings, as there are some noticeable similar pixel-bars in the \newtechniquename (see Fig.~\ref{fig:use-case-reddit} (A)) in which the x-axis is sorted by time. 
% Median and groups 
Next, we apply the implemented reordering strategies to group and highlight similar pixels-bars. 
% Reordering 
The median of each row reorders the y-axis, and we cluster and reorder the embeddings of the x-axis (see Fig.~\ref{fig:use-case-reddit} (B)). 
%  Results 
The first visibly large group of graph embeddings is classified as noise as the embeddings seem to have distinct values in the latent space.
% Next group 
The next groups are clustered together and also have visually similar looking embeddings. 
% Analysis 
We investigated the graphs in groups and between groups by displaying and comparing them in the graph view.
% Interpretation
Thereby, we interpreted and tried to link the embedding characteristics to evolving graph structures.
For example, we noticed that the first group consists of many computer games subreddits (e.g., \textit{pokemongo}) and that the following group contains various political subreddits (e.g., \textit{the\_donald}, \textit{AskTrumpSupporters}, or \textit{politics}). 
% But wait there is more 
We were also able to identify graph structures related to specific time aspects.
For example, the last group (AM) consists of hyperlinks posted only in the morning (between 8-11 am).  
These graphs posted in the morning have specific characteristics (e.g., fewer subreddits) that have been learned by the graph embedding.

% Similarity Search 
\textbf{Searching for Political Events}
% Why 
Next, we search for political events during the 2016 presidential election to identify graph structures with hyperlinks between political subreddits.
% Start 
First, we change the temporal granularity of all embeddings to the duration of 8 hours, which results in approximately 1000 pixel-bars. 
% Selection of election night 
We select the election night of the 8th of November (6 pm - 12 am).
We assume that political subreddits, which posted hyperlinks to other subreddits during the election night, were also active during the election campaign.
% Similarity search ranking 
Afterward, we use the ranking functionality to search in all three graph embeddings for similar embeddings, and we examine the top results. 
% Results 
The top five-nearest neighbors in the three graph embeddings reveal different political events. 
For instance, graph2vec and GL2Vec return the 1st February can be directly linked to Iowa's democratic and republican caucus. 
Other graphs resulting from the similarity search can be related to the democratic nomination of Hillary Clinton (28th July) and Mike Pence being announced as the running mate of Donald Trump (15th July). 
% FGSD
Furthermore, FGSD~\cite{verma2017hunt} ranks the 23rd July high, which can be associated with the Wikileaks email release that revealed a bias of the Democratic Party against Bernie Sanders. 
% Visible 
The publications of Wikileaks are particularly visible in the graph view, as some political subreddits are linked (e.g., \textit{SandersForPresident}, \textit{politics}, \textit{political\_revolution}).
% Summary 
Overall, the use cases describe how \newtechniquename enables identifying temporal changes and states (e.g., political events) and relating the latent space to structural changes in the underlying graph.

%% ------
%% Old stuff 

%% ------
%% Comments 
% \ts{when you can note/describe noticeable bars, could you also automatically detect them? Hence, you could try different embeddings and aggregations to produce maps with patterns automatically. if space, you could describe it in the future work.}

%%%%%%%%%%%%%%%%%%%%%%%%%%%%%%%%%%%%%%%%%%%%%%%%%%%%%%%%%%%%%%%%%%%%%%%%%%%
% Discussion
%%%%%%%%%%%%%%%%%%%%%%%%%%%%%%%%%%%%%%%%%%%%%%%%%%%%%%%%%%%%%%%%%%%%%%%%%%%
\section{Discussion} \label{sec:discussion}
% Intro
The cornerstone of \newtechniquename is the visual analysis of embedded graphs as pixel-based visualization to identify temporal states.
% Steps 
The visualization technique consists of three steps: (1) the multiscale temporal modeling, (2) graph embeddings, (3) the visual analysis of the pixel-based visualization.
% Next 
In the following, we discuss the limitations of \newtechniquename and potential future research directions.

% Parameters of the first two steps 
\textbf{Parameters} 
The (1-2) step has multiple input parameters that profoundly influence the perceived patterns in the pixel-based visualization, such as the latent space size, number of epochs, or the random initialization of the neural network.
% User dependent 
Currently, the parameter choices are set by the user as they depend on many factors, for example, the temporal aggregation depends on the discretization scale of the application domain.
% Advantage 
We consider the usage of various parameters as an advantage of our approach and a possibility for future work to investigate which parameter combinations (e.g., different graph embeddings) can capture distinct temporal changes, such as reoccurring motifs or outlier graphs.

\textbf{Interpretablity} 
The interpretation of the resulting perceivable changes remains challenging due to multiple reasons (see Sec.~\ref{sec:interpretation}), which affects the usability of the approach as the visual encoding is challenging to read.
% Usage 
We consider the interpretation limitation as minor as our approach focuses mainly on highlighting temporal changes.
% Solution 
However, we aim to support the latent space's visual analysis by presenting the underlying embedded graph structures, enabling us to generate new insight into the evolving data and lead to new interpretations.   
% Reordering 
We also offer reordering strategies to examine and interpret neighborhoods and clusters of embeddings in the latent space.
% More context information 
Nevertheless, the extension with further contextual features (e.g., evolving graph metrics) is essential to allow a detailed interpretation and guide users towards interesting patterns.

\textbf{Graph Embeddings}
% Why 
We apply unsupervised graph embeddings to reduce the dimensionality of long sequences of dynamic graphs and automatically learn similarities between large-scale graphs.
% Why better 
In contrast to topological graph metrics (e.g., density), such unsupervised graph embeddings scale to large graphs, do not require any feature engineering, and are domain as well as task agnostic.
% Limitation 
The main limitation of such embeddings is that it remains unclear how many embedding dimensions are required to capture specific structural changes~\cite{salehi2017properties}.
% Future work 
We plan to investigate the required number of dimensions for synthetic temporal patterns and how different input parameters and noise influences the resulting embeddings. 

\textbf{Scalability} 
% Aspects 
For the computational scalability, we consider the graph size ($|V|$ nodes and $|E|$ edges) and the number of time steps $T$. 
% The multiscale temporal modeling 
The (1) step computes supergraphs at multiple levels and requires $O( log(T) \cdot (|V|+|E|) )$ memory and time complexity.
% Speed up 
The computation of the supergraphs can be parallelized to increase the approach's scalability to long sequences of graphs.
% Graph embeddings 
For further reading of runtime complexities of graph embeddings, we refer to the survey of Goyal and Ferrara~\cite{goyal2018graph}, which emphasizes that recent graph embeddings run in $O(|E|)$. 
% Overall runtime 
Therefore, the overall runtime complexity of the approach is $O( log(T) \cdot (|V|+|E|) )$. 
% Precomputation 
We suggest precomputing the embeddings for large scale dynamic graphs on GPU servers, due to the time and memory complexities.
% Main memory 
Once the embeddings have been calculated, they are small enough to fit into the main memory.
% Visual scalability 
Second, the computational efforts affect the interactive visual analysis of the \newtechniquename. 
% Example 
For example, the reordering strategy by clustering scales linearly to the displayed time steps and embedding dimensions.
% Size of the graphs 
Also, the visualization of large scale graphs for the comparison and interpretation in the graph view does not scale to large-scale graphs as the size impairs the node-link diagram's readability. 
% Possible solution 
A possible solution for this issue is to cluster the underlying large-scale graphs and display the identified clusters.
% Challenge 
However, such a clustering makes it challenging to compare graphs as nodes and edges are abstracted into meta-nodes.
% Future work 
Therefore, we plan to examine how different graph embeddings, combined with evolving graph metrics, can be used to compare large-scale graphs.

%% ------
%% Old stuff 

%% ------
%% Comments 

% \ts{Check if you find this idea interesting or possible, maybe it can add to the discussion. But maybe it is too speculative. Would need to think about it.}
% \newtechniquename positions itself between node-link diagrams and matrix representations, in terms of level of detail. 
% NodeTrix \cite{DBLP:journals/corr/abs-0705-0599} combines node-link and matrix representations for scalable large graph analysis. Future work might address how to integrate  \newtechniquename views as an intermediate representation for large graphs. A problem to overcome is how to realize  connections among the representations.

%%%%%%%%%%%%%%%%%%%%%%%%%%%%%%%%%%%%%%%%%%%%%%%%%%%%%%%%%%%%%%%%%%%%%%%%%%%
% CONCLUSION
%%%%%%%%%%%%%%%%%%%%%%%%%%%%%%%%%%%%%%%%%%%%%%%%%%%%%%%%%%%%%%%%%%%%%%%%%%%
\section{Conclusion} \label{sec:conclusion}
% Why 
We presented \newtechniquename, a visualization technique to provide an overview of temporal changes in long and large-scale dynamic graphs. 
% How 
The novel representation consists of the multiscale temporal modeling, unsupervised graph embeddings, and a dense pixel-based visualization to explore the embeddings at different temporal scales.  
% The main idea 
The main idea is to visually analyze the latent space to identify temporal changes in the dynamic graph. 
% Implementation 
The implemented prototype and the use cases show how \newtechniquename can be used to provide insight into evolving graphs and highlight the applicability of the approach to synthetic and real-world dynamic graph data. 
% Generalizability 
Overall, the \newtechniquename is a promising new research direction for dynamic graphs and can be generalized for the visual analysis of unsupervised embedding methods and latent spaces.

%% ------
%% Old stuff 

%% ------
%% Comments 

% The authors wish to thank A, B, and C. This work was supported in part by
% a grant from XYZ.}
\acknowledgments{This work was partly funded by the Deutsche Forschungsgemeinschaft (DFG, German Research Foundation) under Germany's Excellence Strategy - EXC 2117 - 422037984 and the European Union’s Horizon 2020 research and innovation programme under grant agreement No 830892.}

\bibliographystyle{abbrv-doi}

\bibliography{template}
\end{document}